\documentclass{ws-ijmpa}

\usepackage{graphics}
\usepackage{amsmath}
\usepackage{amssymb}
\usepackage{epsfig}
\usepackage{graphics}

\begin{document}

\markboth{Sonia Kabana, Peter Minkowski}
{On the thermal phase structure of QCD at vanishing chemical potentials}

%
\catchline{}{}{}{}{}
%

\title{On the thermal phase structure of QCD at vanishing chemical potentials}

\author{Sonia ~Kabana}
\address{SUBATECH, Ecole des Mines, 4 rue Alfred Kastler, 44307 Nantes, France.}

\author{Peter ~Minkowski}
\address{Albert Einstein Center for Fundamental Physics,
Institute for Theoretical Physics, 
University of Bern, Sidlerstrasse 5,  CH-3012 Bern, Switzerland,
and CERN, Theoretical Physics, CH-1211 Geneva 23, Switzerland.}

\maketitle

\begin{history}
\received{Day Month Year}
\revised{Day Month Year}
\end{history}

\begin{abstract}
The hypothesis is investigated, that the thermal structure of QCD phases at 
and near zero chemical potentials is determined by long range coherence,
inducing the gauge boson pair condensate, and its thermal 
extension, representing a fundamental order parameter. 
A consistent model for thermal behaviour including interactions
is derived in which the
condensate does not produce any latent heat as it vanishes
at the critical temperature inducing a second order phase transition with 
respect
to energy density neglecting eventual numerically small critical exponents.
Localization  and delocalization of color fields are thus separated 
by a unique critical temperature.


\end{abstract}

\ccode{PACS numbers: 12.38.-t,12.38.Mh,05.30.Rt}


\section{Introduction}
\label{intro}

\noindent
The existence and nature of the QCD phase transition is theoretically
accessible to a universal description underlying a thermal system with 
all its general and specific restrictions.
At the moment key theoretical questions concerning the phase structure 
of thermal systems in QCD remain.
Lattice QCD calculations lead to very clear predictions, establishing 
the lack of any phase transition at zero chemical potentials
\cite{WuppBu_fodor,MILC_SU3}.
The validity of this method and its results is not uncontroversial, 
for example see Refs. \cite{10053509,10053672}.

\noindent
There is a longstanding experimental effort to measure characteristic
signatures of the phase transition in collisions of heavy ions at several
facilities at AGS at BNL, SPS at CERN, RHIC at BNL, LHC at CERN and in the
future FAIR at GSI.
The present state of these studies yields significant results, well
interpretable as a strongly interacting partonic matter formed in the
earlier stages of the heavy ion collisions 
\cite{sRHIC,sps}.
These results suggest the existence of a phase transition
in the course of the collision.
Assuming the correctness of this conclusion,
the detailed phase structure and order(s) of assocated transition(s) 
could not be determined conclusively so far .

\noindent
The experimentally measured reactions cover a large range of collision
energies and chemical potentials in thermal analyses.
In particular, with increasing energies the baryochemical potential
in hadron-hadron or heavy ion collisions decreases.
For a comparison reducing these systems to a common baryochemical potential
we refer to \cite{SKPM2001,SK2001}.





\noindent
In this paper we propose to give an "outline in principle"
of the thermal phases of QCD at vanishing chemical potentials
and their relation to the gauge boson pair condensate of QCD.
We proceed to show that the two phases, separated by the critical
temperature $T_{cr}$ at $\mu=0$ can indeed be understood as a
dynamical consequence of uniquely the bosonic pair condensate at nonzero
quark masses, in contrast to supersymmetric models with no
spontaneous breaking of supersymmetry.

\noindent
The two phases are represented {\it approximatively }
by a collection of noninteracting hadrons for $0<T<T_{cr}$ and a collection of
8 gauge bosons and 4 flavours of quarks and antiquarks 
(u,d,s,c \& anti)
for $T> T_{cr}$, in which interactions are modeled.
We use here initially two collections of hadron resonances , 
denoted Ntype=65 and Ntype=26 in the following. 
The Ntype=65 collection, on which most of the subsequent calculations of
thermodynamic state functions rely, 
is an extension of the
smaller Ntype=26 one.  In the former 
$\ \left ( \ \mbox{Ntype=65} \ \right ) \ $
the 4 $\ SU3_{fl} \ $ meson nonets
with $\ J^{\ PC_{n}} \ = \ 0^{++} \ , \ 1^{++} \ , \ 2^{++} \ ; \ 1^{+-} $
in the lowest mass p-wawe $\ q \overline{q}^{'} \ $ configuration 
are added, as well as the lowest mass open charm and anticharm s-wave 
pseudoscalar and vector mesons
with composition $\ q \overline{c} $ and $\ c \overline{q} $ 
and baryons and antibaryons containing one charmed quark and their antiparticles
together with overall s-wave composition $\ q q' c \ $ and 
$\ \overline{q} \overline{q}' \overline{c} $ with 
$\ q , q' \ = \ u,d,s \ $ are added, together
with 3 representative color gauge boson binaries (glueballs) with 
$J^{\ PC} \ = \ 0^{++} , 0^{-+} , 2^{++} $ and two hybrid
nonets of composition $\ q \ g \ \overline{q}' \ $ with 
$\ J^{\ PC_{n}} \ = \ 1^{-+} \ $. 

\noindent
The adopted approximation allows to illustrate the thermal phase properties
in principle.

\section{Energy momentum density tensor and gauge boson condensate}

\noindent
We begin with the local, symmetric and conserved energy momentum density tensor
$ \vartheta_{\ \mu \ \nu} $
existing and renormalized as a basic consequence of QCD .
The above properties imply exact Poincar\'{e} invariance.

\vspace*{-0.4cm}
\begin{equation}
\label{th:1}
\begin{array}{l}
\left \lbrace \ \vartheta_{\ \mu \ \nu} \ = \ \vartheta_{\ \nu \ \mu}
\ \right \rbrace \ ( \ x \ )
\hspace*{0.2cm} ; \hspace*{0.2cm}
\partial^{\ \nu} \ \vartheta_{\ \mu \ \nu} \ = \ 0
\end{array}
\vspace*{-0.1cm}
\end{equation}

\noindent
The central defining quantities, of the existence and order of the
phase transition, are pressure, energy density and entropy density
for zero chemical potentials.
The {\it local} operator form of the trace anomaly in QCD reads:


\vspace*{-0.4cm}
\begin{equation}
\label{th:2}
\begin{array}{l}
\vartheta^{\ \mu}_{\hspace*{0.3cm} \mu}
 = \sum_{ f} m_{ f} \ S_{ \dot{f}  f} \ ( x  )  
+ \ \delta_{ 0} \ ( x )  
\vspace*{0.1cm} \\
\delta_{\ 0}  = \begin{array}[t]{l} \left ( - 2 \beta ( g  ) 
\ / \ g^{ 3}  \right ) 
 \left \lbrack \ - \frac{1}{4}  (:) \ F_{\mu \nu}^{ A} \ (  x  )
\ F^{ \mu  \nu \ A} \ (  x  )  (:)
  \right \rbrack_{s. d.} \end{array}
\vspace*{0.1cm} \\
\hspace*{0.0cm} = - 2 b_0 
 \left \lbrack \frac{1}{4}  (:) F_{ \mu \nu}^{ A} \ (  x  )
\ F^{ \mu \nu \ A} \ (  x  ) (:)
  \right \rbrack
\vspace*{-0.4cm}
\end{array}
\vspace*{0.35cm}
\end{equation}

\noindent
In Eq. \ref{th:2} the subscript
${s.d.}$ stands for "renormalization scale dependent", 
while $b_0 = 9/ (16 \pi^2) $.
All quantities are defined to be renormalized and renormalization
group invariant, except those with the subscript ${s.d.}$
$ \beta (g)$ denotes the (Callan-Symanzik-) rescaling function, where $g$ is
the coupling constant. 
$ \ \frac{1}{4} \ (:) \ F_{\ \mu \ \nu}^{\ A} \ F^{\ \mu \ \nu \ A} \ (:) $
is the {\it local} scalar density -- to be called
{\it local} gauge boson bilinear --
composed bilinearly of field strength tensors $F_{\ \mu \ \nu}^{\ A}$ -- the
latter including {\it multiplicatively} the gauge coupling constant 
in their definition, relative to the perturbative field strength normalization.

\noindent
$A$ denotes the color component within the adjoint representation of
$SU(3)_c$. (:)(:) stands for a {\it suitable} normal ordering.
$m_{ f}$ denotes the mass of quark flavour f :
$S_{\ \dot{f} f} $ denotes the {\it local} scalar density 
of antiquark $\dot{f} $
and quark $f$ flavors : \\
$S_{\dot{f} f} ( x ) = 
(:) \ \overline{q}_{ \dot{f} \dot{c} s} ( x ) 
 {q}_{f c }^{ s} ( x ) \ (:)$, 
where s denotes the spin.






\noindent
The focus is to consider vacuum expected values of the {\it local}
operators in Eq. \ref{th:2}, which by translation invariance are independent of
the position $x$, suppressed in the following 

\vspace*{-0.3cm}
\begin{equation}
\label{th:3}
\begin{array}{l}
\eta^{ \mu \nu} \left \langle \Omega \right | 
 \vartheta_{ \mu \nu} (x) \ \left | \Omega \right \rangle 
 = 
- 2 b_0 
\left \langle \Omega \right |
\frac{1}{4}  (:)  F_{ \mu  \nu}^{ A} \ F^{ \mu \nu A} (x) (:)
\left | \Omega \right \rangle 
\vspace*{0.1cm} \\
+ \ m_f \ \left \langle  \Omega \ \right |
 S_{\dot{f} f} \ (  x  )
 \left |  \Omega  \right \rangle
\end{array}
\end{equation}


\noindent
$\left \langle \ \Omega \ \right | (:) \frac{1}{4}  F_{ \mu \nu}^{ A} 
\ F^{ \mu \nu \ A}  (:)  \left | \ \Omega \ \right \rangle  = 
 {\cal{B}}^{ 2} $ shall be called the gauge boson pair condensate ,
abbreviated by $\ {\cal{B}}^{\ 2} \ $ .

\noindent
$ \left \langle \ \Omega \ \right | \ S_{\ \dot{f} \ f}
\ \left | \ \Omega \ \right  \rangle \ \neq \ 0$
induces spontaneous chiral symmetry breaking and is generally called 
quark condensate.




\noindent
In connection with {\it normal ordering ambiguities} it is important to admit
in the precise form of the energy momentum tensor a nontrivial vacuum
expected value , which as a consequence of {\it exact} Poincar\'{e} 
invariance must be of the form

\vspace*{-0.4cm}
\begin{equation}
\label{th:4}
\begin{array}{l}
\left \langle \ \Omega \ \right | \ \vartheta_{ \mu  \nu} \ ( x )
\ \left | \ \Omega \ \right \rangle  =
 - \eta_{ \mu  \nu} \ p_{ vac}
\vspace*{0.1cm} \\ 
\left \lbrace \begin{array}{c}
\eta_{ \mu  \nu}  =  diag \ \left (  1  ,  -1  ,  -1  ,  -1
 \right )
\hspace*{0.1cm} ; \hspace*{0.1cm}
p_{ vac} = - \rho_{ vac}
\end{array} \right \rbrace 
\vspace*{0.1cm} \\
\mbox{independent of $x$}
\hspace*{0.1cm}  \rightarrow
\vspace*{0.1cm} \\
\Delta  \vartheta_{ \mu  \nu} \ (  x  )  = 
 \vartheta_{ \mu  \nu} \ (  x  )  -  
 \left \langle  \Omega  \right |  \vartheta_{ \mu  \nu} \ (  x  )
 \left |  \Omega  \right \rangle 
\hspace*{0.1cm} \left |  \Omega  \right \rangle
\ \left \langle  \Omega  \right | 
\vspace*{0.1cm} \\ 
\mbox{with} \hspace*{0.2cm} \partial^{\ \nu} \ \Delta \ \vartheta_{\ \mu \ \nu}
\ ( \ x \ ) \ = \ 0 
\hspace*{0.1cm} ; \hspace*{0.1cm} 
\left \langle \ \Omega \ \right | \ \Delta  \vartheta_{ \mu  \nu}
\ (  x  ) \ \left | \ \Omega \ \right \rangle =  0 
\end{array}
\end{equation}

\noindent
In Eq. \ref{th:4}
$\ P_{\ \Omega} \ = \ \left | \ \Omega \ \right \rangle
\ \left \langle \ \Omega \ \right | \ $ denotes the projector on the ground
state .

\noindent
From the two local, {\it conserved} tensors in Eq. \ref{th:4}
only $ \Delta \vartheta_{ \mu \nu} ( x )$ with vanishing vacuum
expected value is acceptable as  representing the conserved 4 momentum
{\it operators} and their densities yielding the integral form

\vspace*{-0.3cm}
\begin{equation}
\label{th:5}
\begin{array}{l}
\widehat{P}_{ \mu} \ = \ {\displaystyle{\int}\vspace*{-0.0cm}}\vspace*{-0.0cm}_{ t} \ d^{ 3} 
\ x
\ \Delta \vartheta_{ \mu  0} \ (  t  ,  \vec{x} )
\vspace*{-0.0cm}
\end{array}
\vspace*{-0.2cm}
\end{equation}
\noindent

We use here throughout strictly thermal , 'extension in phase space'
associated potentials, depending in subtle ways on vacuum condensates.
To these potentials no vacuum associated spontaneous parameters like 
$ p_{ vac} \ = \ - \rho_{ vac}  $, defined in Eq. \ref{th:4},
contribute in a direct way, dominating in the limit 
$\ T \ \rightarrow \ 0 \ ; \ \mu_{\ \alpha} \ \equiv \ 0 \ $.



\noindent
From Eqs. \ref{th:3} and \ref{th:4}
we obtain the relation and estimates 'pour fixer les id\'{e}es'

\vspace*{-0.4cm}
\begin{equation}
\label{th:6}
\begin{array}{lll}
p_{ vac} & = & \begin{array}{c}
9
\vspace*{0.15cm} \\
\hline  \vspace*{-0.3cm} \\
32 \ \pi^{\ 2}
\end{array}
\ {\cal{B}}^{\ 2}
\ + \ \frac{1}{4} \ \Lambda 
 =  \left \lbrace
\begin{array}{l}
0.00302 \ \mbox{GeV}^{\ 4}
\vspace*{0.0cm} \\
0.00658 \ \mbox{GeV}^{\ 4}
\end{array} \right . 
\vspace*{0.1cm} \\
{\cal{B}}^{\ 2} & = 
& \left \lbrace
\begin{array}{l}
0.125 \ \mbox{GeV}^{\ 4} \ \mbox{\cite{SVZ1979}}
\vspace*{0.0cm} \\
0.250 \ \mbox{GeV}^{\ 4} \ \mbox{\cite{Nari1998}}
\end{array} \right .
\vspace*{0.2cm} \\
\Lambda & = & - \sum_{ f} \ m_{ f}
 \left \langle \Omega \right | \ S_{ \dot{f}  f}  
\ \left |  \Omega \right  \rangle 
\vspace*{0.1cm} \\
& \sim & f_{\ \pi}^{\ 2} \ ( \ \frac{1}{2} \ m_{\ \pi}^{\ 2} \ + 
\ m_{\ K}^{\ 2} \ ) 
 =  0.00217 \ \mbox{GeV}^{\ 4}

\vspace*{-0.6cm}
\end{array}
\vspace*{0.3cm}
\end{equation}








\section{Construction of a thermal model including interactions}

\noindent
We follow the strategy layed out in Refs. \cite{SKPM2001,SK2001} taking into 
account
the modifications described above, distinguishing 
two {\it eventual} phases

\vspace*{-0.2cm}
\begin{description}
\item 1) the hadronic (h)-phase ,
with color localized within stable hadrons 
and selected hadron resonances corresponding to the two collections
Ntype=65$\ \subset \ $26.
Thermal potentials of the 
(h)-phase are approximated by those of free
hadrons, neglecting decay widths, as described in Refs.
\cite{SKPM2001,SK2001}.  

\vspace*{0.2cm}
\item 2) the quark-antiquark-gauge boson (qg)-phase,
wherein thermal potentials are related but not equal to those of free quarks
and antiquarks, restricted to the flavors u,d,s and c,
and 8 gauge bosons pertaining to the gauge group 
$SU3_{ c} $.
Next we describe the modeling of interactions in the (qg)-phase, which
deviates from noninteracting constituents assumed in Refs. 
\cite{SKPM2001,SK2001} .

\end{description}

\noindent
We introduce for later use for the (qg)-phase , 
the Gibbs density $ \ g^{\ (0)}_{\ qg} \ $
and energy density $\ \varrho_{\ e \ qg}^{\ (0)} \ $ 
pertaining to noninteracting tricolored quarks , antiquarks with flavors \\
u , d , s , c and eightfold colored gauge bosons 

\vspace*{-0.5cm}
\begin{equation}
\label{th:7}
\begin{array}{l}
g_{ qg}^{ (0)} \ (  T  )  =  \sum_{ \alpha_{qg}}
w_{ \alpha_{ qg}} \ \left (  1  /  (  2 \pi^{ 2}  )  \right ) 
\ {\displaystyle{\int}}_{ m_{ \alpha_{qg}}}^{ \infty}
\ l \ E \ p \ d \ E
\vspace*{0.1cm} \\
w_{ \alpha_{qg}}  = 
 \left (  2  spin_{ \alpha_{qg}}  +  1  \right )
 \left \lbrace \begin{array}{l}
3 \ \mbox{for} \ q  ,  \overline{q}
\\ 8 \ \mbox{for} \ g 
\end{array} \right .
= \ \left \lbrace \begin{array}{l}
\ 6 \ \mbox{for} \ q  ,  \overline{q}
\\ 16 \ \mbox{for} \ g
\end{array} \right .
\vspace*{0.2cm} \\
\beta  \equiv  1  /  T
\hspace*{0.05cm} ; \hspace*{0.05cm}
l  =  \mp  \log  \left \lbrack  1  \mp  \exp \ ( - \beta  E  ) 
 \right \rbrack
\vspace*{0.1cm} \\
\varrho_{ e \ qg}^{ (0)} \ ( T  )  = 
 -  \left (  d  /  d  \beta  \right ) \ g_{ qg}^{ (0)} \ (  T  ) 
\ = \ T^{\ 2} \ \left (  d  /  d  T  \right )
\ g_{\ qg}^{ (0)} \ (  T  )
\vspace*{-1.1cm}
\end{array}
\vspace*{1.0cm}
\end{equation}

\noindent
In Eq. \ref{th:7} the index $\ \alpha_{qg} \ $ runs over the different
constituents of the (qg) phase, while $\ w_{\ \alpha_{qg}} \ $
denotes the multiplicity beyond momentum phase space associated with the
constituent $\ \alpha_{qg} \ $. The sign ( $\mp$ ) in the expression for 
$\ l \ $ is - for bosons and + for fermions .

\noindent
We choose the following masses for
quark flavors u, d, s, c 

\vspace*{-0.2cm}
\begin{equation}
\label{th:8}
\begin{array}{l}
\left \lbrack \ m_{ u} = 0.00525  ,  m_{ d}  =  0.00875 \right . \\ 
\left . \hspace*{0.2cm} m_{ s}  =  0.175 ,
 m_{ c}  =  1.27 \ \right \rbrack \ \mbox{GeV}
\vspace*{-0.1cm}
\end{array}
\vspace*{0.0cm}
\end{equation}

%
%

\noindent
The absolute masses of the u,d,s light flavors as well as  their ratios 
$\ m_{\ u} \ : \ m_{\ d} \ : \ m_{\ s} \ = \ 3 \ : \ 5 \ : \ 100 \ $,
\cite{minkzepeda,dominguez,narison} ,
used here play no decisive role in the present derivations, within generous
ranges of $\pm$ 20 \% .

\noindent
The inclusion of the charmed quark ser\-ves the purpose to check whether 
it has any significant influence on the thermal parameters in the region of
$\ T_{\ cr} \ \sim 0.2 \ \mbox{GeV} \ $, wich turns out to be in the
few percent range .


\noindent
We proceed to modify the free quark antiquark gauge boson 
(qg-) parametri\-zation of the Gibbs potential and the energy density, which for
$\ \mu_{\ \alpha} \ = \ 0 \ $ must obey the exact relation

\vspace*{-0.3cm}
\begin{equation}
\label{th:9}
\begin{array}{l}
g_{ qg}  =  g_{ qg} \ (  T  \equiv \beta^{\ -1} \ )
\hspace*{0.1cm} ; \hspace*{0.2cm}
-  \left (  d  /  d  \beta  \right ) \ g_{ qg} \ (  T  )
 =  \varrho_{ e \ qg} \ (  T  )
\vspace*{0.0cm} \\
\mbox{and} \hspace*{0.2cm}
g_{\ qg} \ \leftrightarrow \ g_{\ h} 
\hspace*{0.2cm} , \hspace*{0.2cm}
\varrho_{ e \ qg}  \leftrightarrow  \varrho_{ e \ h} 
\vspace*{-0.3cm}
\end{array}
\vspace*{0.2cm}
\end{equation}

\noindent
The Gibbs- and energy-densities in the hadron phase are constructed 
from the expressions analogous to the ones given in Eq. \ref{th:7} ,
where the index $\alpha_{qg} \ \rightarrow \ \alpha_{h}$ 
runs over a suitable choice of hadrons and hadron resonances as defined
in Refs. \cite{SKPM2001,SK2001} with real masses and neglecting interactions
among these states .

\noindent
The ensuing parametrization of interactions is understood as representing the
phase structure in principle and not in numerical detail.
The modeling of interactions in the qg-phase is performed
setting  two parameters $\ k \ , \ \Delta \ g \ $ , independent of
temperature, as approximately parametrizing the interaction in the
deconfined phase in a limited region of $\ T \ \geq \ T_{\ cr} \ $

\vspace*{-0.3cm}
\begin{equation}
\label{th:10}
\begin{array}{l}
\begin{array}{lll}
\varrho_{e \ qg} (  T  ;  k  ) & = & k \ \varrho_{e \ qg}^{ (0)} \ (  T  )
\vspace*{0.0cm} \\
g_{ qg} ( T ;  k  ,  \Delta g  )
& = & k \ g_{ qg}^{\ (0)} (  T  )  -  \Delta \ g
\end{array}
\vspace*{-0.5cm}
\end{array}
\vspace*{0.5cm}
\end{equation}

\noindent
The parameter $\ 0 \ < \ k \ < \ 1 \ $ is taking into account the reduction
of Gibbs density or pressure relative to the noninteracting (Stefan-Boltzmann)
limit , noted in perturbative QCD calculations of thermal parameters 
for $T \ \simeq \ T_{\ cr} \ $ of interest here \cite{Reb2007} , while
the second parameter $\ \Delta \ g \ $ arises as integration constant
from the differential equation ( Eq. \ref{th:9} ) , which is clearly satisfied
for arbitrary values of $\ \left ( \ k \ , \ \Delta \ g \ \right ) \ $.

\noindent
We proceed in two steps to map out the structure of the phase transition
\vspace*{-0.3cm}

\begin{description}
\item  I : the condition determining
$ T_{ cr}  \leftrightarrow  k $

\noindent
The equality of the energy densities -- in the hadron phase
$\ \varrho_{e \ had} \ \mbox{for} \ T \ \leq \ T_{\ cr} \ $ as outlined in
Refs. \cite{SKPM2001,SK2001} and in the qg-phase as defined in Eq. \ref{th:10}
$\ \varrho_{\ e \ qg} \ \mbox{for} \ T \ \geq \ T_{\ cr} \ $ 
determine the critical temperature 

\vspace*{-0.7cm}
\begin{equation}
\label{th:11}
\begin{array}{l}
\varrho_{ e \ had} \ (  T  )  =  \varrho_{ e \ qg} \ (  T  ;  k  )
\hspace*{0.2cm} \leftrightarrow \hspace*{0.2cm}
T  =  T_{ cr} \ (  k  )
\end{array} 
\vspace*{-0.2cm}
\end{equation}

The matching ( Eq. \ref{th:11} ) is further restricted to yield the value 

\vspace*{-0.5cm}
\begin{equation}
\label{th:12}
\begin{array}{l}
T_{ cr}  \sim  0.2 \ \mbox{GeV}
\hspace*{0.1cm} \leftrightarrow \hspace*{0.1cm}
k  \sim  0.365
\end{array}
\vspace*{-0.1cm}
\end{equation}

\noindent
in accordance with the estimate of one of us \cite{PMBech1988} .

\item II : the condition avoiding singular behaviour of pressure gradient

\noindent
This condition implies

\vspace*{-0.8cm}
\begin{equation}
\label{th:13}
\begin{array}{l}
g_{ h} \ (  T_{ cr}  )  = 
 g_{ qg} \ (  T_{ cr}  ;  k  ,  \Delta  g  ) 
\hspace*{0.1cm} \leftrightarrow \hspace*{0.1cm}
\Delta g  =  \Delta  g \ (  T_{ cr} \ (  k  )  )
\vspace*{-0.0cm}
\end{array}
\vspace*{+0.0cm}
\end{equation}

\noindent 
and determines $\ \Delta g \ $

\vspace*{-0.3cm}
\begin{equation}
\label{th:14}
\begin{array}{l}
\Delta  g  \sim  
0.0062  \hspace*{0.1cm} \mbox{GeV}^{\ 3}
= \ 0.775 \hspace*{0.1cm} \mbox{fm}^{\ -3}
\vspace*{-0.3cm}
\end{array}
\vspace*{0.1cm}
\end{equation}

\end{description}

\section{Results and discussion}

\noindent
We begin with a disussion with a comparison of the two colletions of hadron
resonances Ntype=65 and Ntype=26 whence thermal state functions are
extrapolates beyond the
temperature range $\ 130 \ \mbox{MeV} \ \leq \ T \ \leq \ 170 \ \mbox{MeV} \ $
obtained in studies of chemical equilibrium of hadron yield ratios.

\noindent
It is expected that approaching a phase transition will require 
interactions on the hadronic side as well as on the quark-gluon side.
The mere extension of the collection of hadron resonances also beyond Ntype=65
here , without introduction of interactions among the hadron resonances is
unrealistic. If there is no phase transition at all, this does not change 
the mentioned expectation.

\noindent
This is illustrated in Figs. 1 and 2 , in which the two noninteracting 
collections of hadron resonances, Ntype=65 and 26, are compared with respect to the quantities 
dscale = $\ \left ( \ \varrho_{e} \ - \ 3 \ p \ \right ) \ / \ T^{4} \ $
and pressure / \ $T^{\ 4}$ respectively .



\vspace*{-0.0cm}
\begin{center}
\hspace*{0.0cm}
\begin{figure}[htb]
\vskip -0.8cm
\hskip -0.0cm
\includegraphics[angle=-90,width=12.0cm]{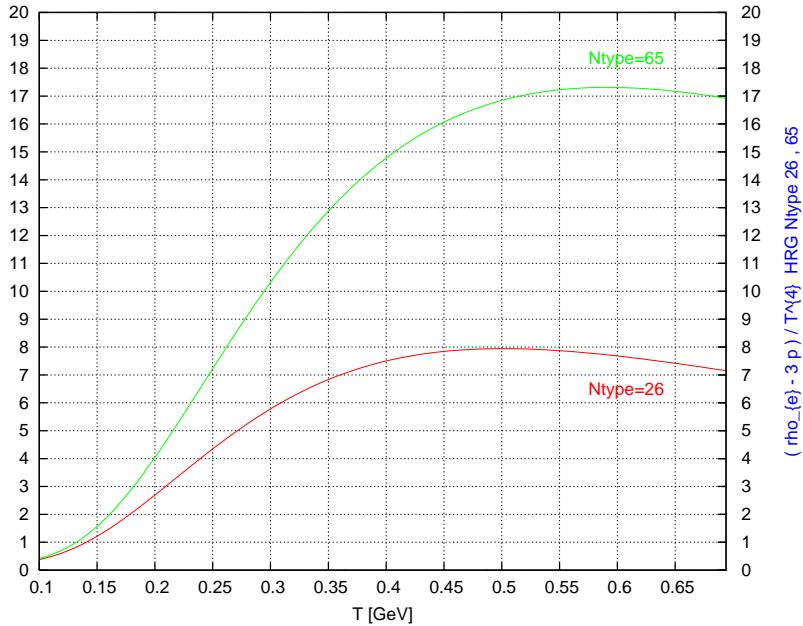}
\vskip -0.0cm
\caption{The thermal quantity representing the trace anomaly -- dscale $\ ( \ T \ ) \ $ =
$\ \left ( \ \varrho_{e} \ - \ 3 \ p \ \right ) \ / \ T^{4} \ $ is shown for
$\ 0.1 \ \mbox{GeV} \ \leq \ T \ \leq \ 0.7 \ \mbox{GeV} \ $. The
comparison of HRG collections Ntype=65 and 26 shows the sensitive temperature
regions beyond temperatures where chemical freeze out takes place  
-- $\ T \ \sim \ 150 \ - \ 170 \ \mbox{MeV} \ $ for Pb - Pb collisions
at SPS and Au - Au collisions at RHIC.
}
\label{fig1}
%
%
\vspace*{-0.6cm}
\end{figure}
\end{center}

\newpage

\vspace*{-0.0cm}
\begin{center}
\hspace*{0.0cm}
\begin{figure}[htb]
\vskip -0.7cm
\hskip -0.0cm
\includegraphics[angle=-90,width=12.0cm]{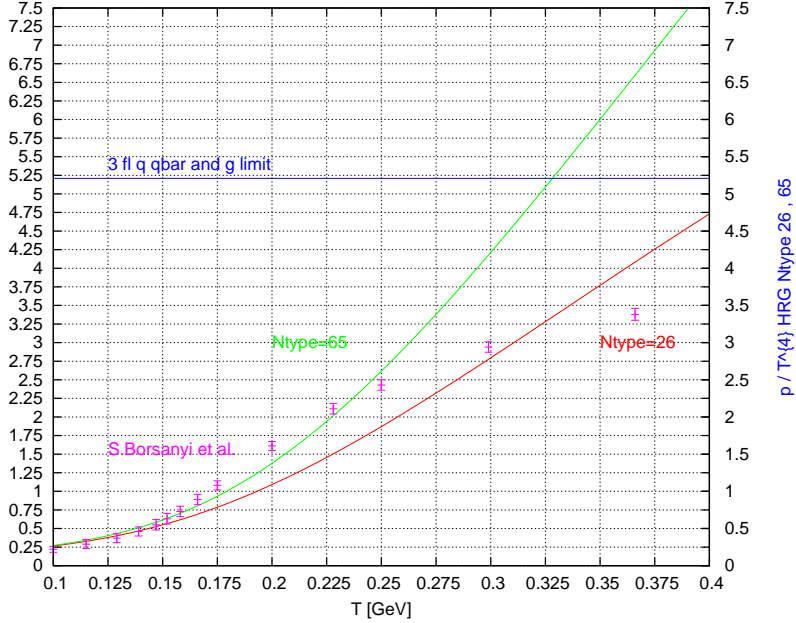}
\vskip -0.0cm
\caption{ Pressure $\ / \ T^{4} \ $ versus T
in the range $\ 0.1 \ \mbox{GeV} \ \leq \ T \ \leq \ 0.4 \ \mbox{GeV} \ $, 
for the two collections Ntype=65 and 26 is compared with the results
of 
lattice QCD by S. Borsanyi et al. 
\protect \cite{Fodoretal} .
}
\label{fig2}
\end{figure}
\vspace*{-0.3cm}
\end{center}

\noindent
As a consequence of the fair agreement of the pressure $\ / \ T^{4} \ $
as obtained from the noninteracting hadron resonance collection Ntype=65 with
the lattice simulation results of S. Borsanyi et al. 
\cite{WuppBu_fodor,Fodoretal} , as shown
in Fig. 2 , we choose this collection in all subsequent calculations
representing the hadron phase for 
$\ T \ \leq \ T_{ cr} \ \simeq \ 200 \ \mbox{MeV} \ $.

\noindent
We mention the
fact, that collections of hadron resonances used in recent analyses of
especially ratios od hadron yields as observed in Au-Au collisions at RHIC
include all resonances reported by the PDG \cite{PDG} up to a mass of
2 GeV or 2.5 GeV , which are larger than the one defined Ntype=65 here.
In Ref. \cite{WuppBu_fodor} the agreement up to $\ T \ \sim \ 170 \ \mbox{MeV} 
\ $
of all thermal state functions with the HRG quantities of these large
collections is very satisfactory .

\noindent
The restriction to noninteracting hadron resonances nevertheless has a limited
validity particularly at temperatures above  chemical freeze out as 
$\ T \ \sim \ 200 \ \mbox{MeV} \ $.


\newpage

\noindent
The result of steps I and II above are illustrated in Figs. 3 - 7
below . For a comparison with thermodynamic analyses in lattice QCD we
refer to Refs. \cite{FKar2009}, \cite{Baza2009} and \cite{WuppBu_fodor}. 
The existence of 
two crossing points, one each for energy density and pressure, coincident
in temperature at acceptable values of all parameters near 
$\ T_{cr} \ \simeq \ 0.2 \ \mbox{GeV} \ $ is nontrivial , shown in Fig. 3
below


\vspace*{-0.2cm}
\begin{center}
\hspace*{0.0cm}
\begin{figure}[htb]
\vskip -0.0cm
\hskip -0.0cm
\includegraphics[angle=-90,width=9.0cm]{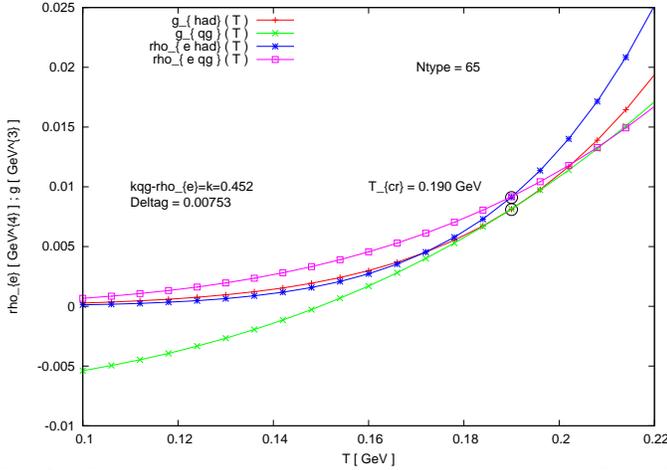}
\vskip -0.2cm
\caption{The two thermal quantities energy density $\ \varrho_{\ e} \ $ 
and Gibbs density $\ g \ = \ p \ / \ T \ $ in units of $\ \mbox{GeV}^{\ 4} \ $
and $\ \mbox{GeV}^{\ 4} \ $  respectively are decomposed into the hadronic (had)
( valid for $\ T \ \leq \ T_{ cr} \ $)
and quark-gluon (qg) parts ( valid for $\ T \ \geq \ T_{ cr} \ $) each :
$\ \varrho_{\ e \ had} \ $ : blue curve ; $\ \varrho_{\ e \ qg} \ $ : purple
curve and $\ g_{\ had} \ $ : red curve ; $\ g_{\ qg} \ $ : green curve .
The two small circles mark the two equalities 
$\ \varrho_{\ e \ had} \ = \ \varrho_{\ e \ qg} \ $ and 
$\ g _{\ had} \ = \ g_{\ qg} \ $ coincident in temperature 
$\ T \ = \ T_{\ c} \ \left ( \ \equiv \ T_{ cr} \ \right ) \ $.
}
\label{fig3}
\end{figure}
\end{center}

\noindent
In Figs. 4 - 6 the modifications of $\ \varrho_{e \ qg} / T^{4} \ $, 
$\ p_{\ qg} / T^{4} \ $ and 
$\ \mbox{dscale} \ = \ \left ( \ \varrho_{\ e \ qg} \ - \ 3 \ p_{\ qg} 
\ \right ) \ / \ T^{4} \ $
introduce a subleading critical exponent $\nu \ ; \ \nu \ = \ 0.975 \ $
used here throughout,
in the vicinity of $T = T_{cr}$ allowing the free quark-gluon limits to be
reached for $T \rightarrow \infty$ . The modified quantities are
defined as


\vspace*{-0.2cm}
\begin{equation}
\label{th:14b}
\begin{array}{l}
\left \lbrace \begin{array}{l}
\varrho_{e \ qg}^{\ \nu}
\vspace*{0.0cm} \\
p_{\ qg}^{\ \nu}
\end{array} \right \rbrace
\ = 
\ \left \lbrace \begin{array}{l}
fmod \ ( \ \nu \ , \ T \ / \ T_{\ c} \ ) \ \varrho_{e \ qg}
\ + \ fmod1 \ ( \ \nu \ , \ T \ / \ T_{\ c} \ ) \ p_{\ qg}
\vspace*{0.0cm} \\
fmod \ ( \ \nu \ , \ T \ / \ T_{\ c} \ ) \ p_{\ qg}
\end{array} \right \rbrace
\vspace*{0.1cm} \\
fmod \ ( \ \nu \ , \ T \ / \ T_{cr} \ ) \ = \ 1 + 
\ \left | \ 1 - T_{cr} \ / \ T \ \right |^{\ 2 \nu}
\ \left ( \ 1 \ / \ k - 1 \ \right )
\vspace*{0.1cm} \\
fmod1 \ ( \ \nu \ , \ T \ / \ T_{\ c}) \ = \ T \ ( \ d \ / \ d \ T \ ) 
\ fmod \ ( \ \nu \ , \ \ T \ / \ T_{\ c} \ )
\end{array}
\end{equation}

\noindent
Assuming that all thermodynamic potentials are analytic functions of $\ T \ $
in a sufficiently large neighbourhood of $\ T_{\ cr} \ $ , 
with the transition arising from appropriate conditional choices below and
above $\ T_{\ cr} \ $, as is the case for the 
parametrizations adopted here, the orders of the so deduced phase transition 
( for $\ \mu_{\ \alpha} \ = \ 0 \ $ ) become

\vspace*{-0.7cm}
\begin{equation}
\label{th:15}
$$\begin{tabular}{ccl}
order & \begin{tabular}{c} with 
           respect to \end{tabular} & \hspace*{-0.3cm} 
	   quantity 
\vspace*{0.1cm} \\ \hline \vspace*{-0.3cm} \\
2. & $\leftrightarrow$ & \hspace*{-0.3cm} energy density 
\vspace*{-0.0cm} \\
3. & $\leftrightarrow$ & \hspace*{-0.3cm} pressure
\vspace*{-0.0cm} \\
1. & $\leftrightarrow$ & \hspace*{-0.3cm} sound velocity
\vspace*{-0.7cm}
\end{tabular}$$
\vspace*{-0.0cm}
\end{equation}



\noindent
We note the behaviour of the square velocity of sound. 
Steps I and II imply that 
the sound velocity (square) exhibits a 1. order transition at
$\ T \ = \ T_{\ cr} \ $ .

\noindent
Let the squares of the sound velocity in the hadron phase 
( qg-phase ), always for
vanishing chemical potentials, be denoted respectively

\vspace*{-0.3cm}
\begin{equation}
\label{th:16}
\begin{array}{l}
v_{ s \ had \ (qg)}^{ 2} \ (  T  )  =
 \begin{array}{c}
\dot{p}_{ had \ (qg)}
\vspace*{0.2cm} \\
\hline  \vspace*{-0.4cm} \\
\dot{\varrho}_{ e \ had \ (qg)}
\end{array} 
\hspace*{0.2cm} ; \hspace*{0.2cm}
\dot{} \ = \ d \ / \ d \ T 
\end{array}
\end{equation}

\noindent
It follows that the sound velocity (square)  jumps from a lower
value in the hadron phase to a higher value in the qg-phase .



\noindent
For the qg-phase we have the relations

\vspace*{-0.5cm}
\begin{equation}
\label{th:17}
\begin{array}{l}
p_{ qg} \ = \ T \ \left ( k \ g_{ qg}^{\ (0)}  -  \Delta  g  \right )
\hspace*{0.2cm} , \hspace*{0.2cm}
T \ g_{ qg}^{ (0)}  =  p_{ gq}^{\ (0)} \ (  T  ) 
\vspace*{0.1cm} \\
\dot{p}_{\ qg}^{ (0)}  =  \varrho_{ s}^{ (0)} \ (  T  )
\hspace*{0.2cm} : \hspace*{0.2cm}
\mbox{{\tiny \begin{tabular}{l} entropy density \\
for noninteracting \\
qg-phase \end{tabular}}} 
\vspace*{0.2cm} \\
\longrightarrow \hspace*{0.2cm}
v^{ 2}_{ qg}  = 
 \begin{array}{c}
k \ \dot{p}_{ qg}^{ (0)}  -  \Delta  g
\vspace*{0.2cm} \\
\hline  \vspace*{-0.4cm} \\
k \ \dot{\varrho}_{ e \ qg}^{ (0)} 
\end{array} 
 =  \left ( \ v^{ 2}_{ qg} \ \right )^{ (0)}  -
\ \begin{array}{c}
\Delta  g
\vspace*{0.2cm} \\
\hline  \vspace*{-0.4cm} \\
k \ \dot{\varrho}_{ e \ qg}^{ (0)}
\end{array}
\vspace*{-0.3cm}
\end{array}
\vspace*{0.1cm}
\end{equation}

\noindent
All quantities with superscript $\ ^{(0)} \ $ refer to the noninteracting
qg-phase . The introduction of interactions leads
to a reduction of the qg-sound velocity square proportional to $\ \Delta \ g \
$ . This is however
a small correction for {\it all} $\ T \ > \ T_{\ cr} \ $ as shown in Fig. 4,
also illustrating the 1. order transition of $\ v^{\ 2}_{\ s} \ $ . In the
qg-phase
the sound velocity square approaches the limiting value $\ \frac{1}{3} \ $ and
is just after the transition
already quite near this value.
\noindent
The behaviour of $\ v_{\ s}^{\ 2} \ $ is shown in Fig. 7 .



\vspace*{-0.2cm}
\begin{center}
\hspace*{0.0cm}
\begin{figure}[htb]
\vskip -0.4cm
\hskip -0.0cm
\includegraphics[angle=0,width=10.5cm]{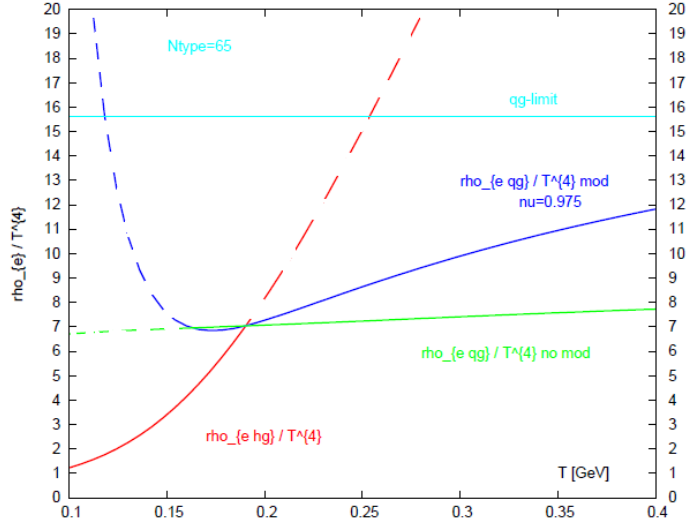}
\vskip -0.2cm
\caption{
The ( dominantly ) second order nature is displayed of the transition
-- unmodified and modified on the quark-gluon-side by subleading critical
exponents --
of the quantity
$\ \varrho_{\ e} \ ( \ T \ ) \ / \ T^{\ 4} \ $ , piecewise described by
$\ \varrho_{\ e \ hg(-65)} \ / \ T^{\ 4} \ $ for $\ T \ \leq \ T_{ cr} \ $
from the HRG side and $\ \rho_{\ e \ qg \ no \ mod \ (mod)} \ / \ T^{\ 4} \ $
for $\ T \ \geq \ T_{ cr} \ $ from the quark-gluon side as defined in Eq.
\protect \ref{th:14b}.
The three curves are represented with dashed lines outside their range
of validity.
}
\label{fig4}
\vspace*{+0.10cm}
\end{figure}
\end{center}


\vspace*{-0.2cm}
\begin{center}
\hspace*{0.0cm}
\begin{figure}[htb]
\vskip -0.75cm
\hskip -0.0cm
\includegraphics[angle=0,width=10.5cm]{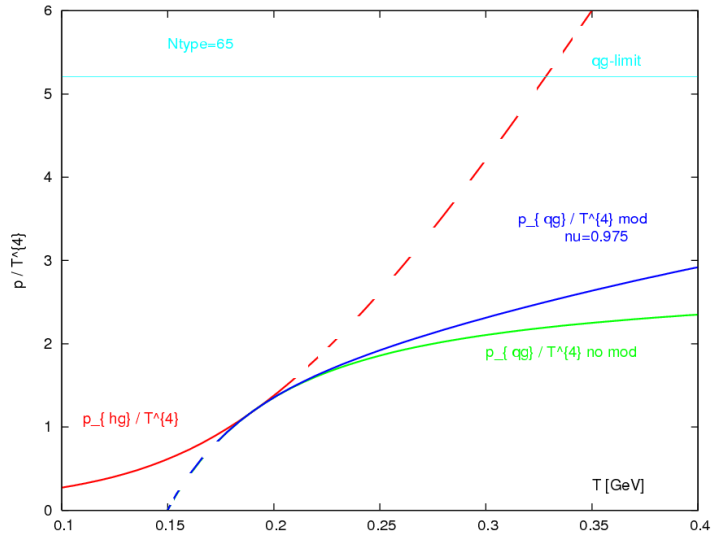}
\vskip -0.3cm
\caption{
The ( dominantly ) third order nature is displayed of the transition under the 
same conditions as underlying Fig. 4 for the quantity 
$ \ p \ / \ T^{\ 4} \ $ , piecewise described by
$\ p_{\ e \ hg(-65)} \ / \ T^{\ 4} \ $ for $\ T \ \leq \ T_{ cr} \ $
from the HRG side and $ \ p_{\ qg \ no \ mod \ (mod)} \ / \ T^{\ 4} \ $ for 
$\ T \ \geq \ T_{ cr} \ $ from the quark-gluon side as defined in Eq.
\protect \ref{th:14b} . 
The three curves are represented with dashed lines outside their range
of validity.
}
\label{fig5}
\end{figure}
\end{center}


\vspace*{-0.2cm}
\begin{center}
\hspace*{0.0cm}
\begin{figure}[htb]
\vskip -0.4cm
\hskip 0.0cm
\includegraphics[angle=0,width=12.0cm]{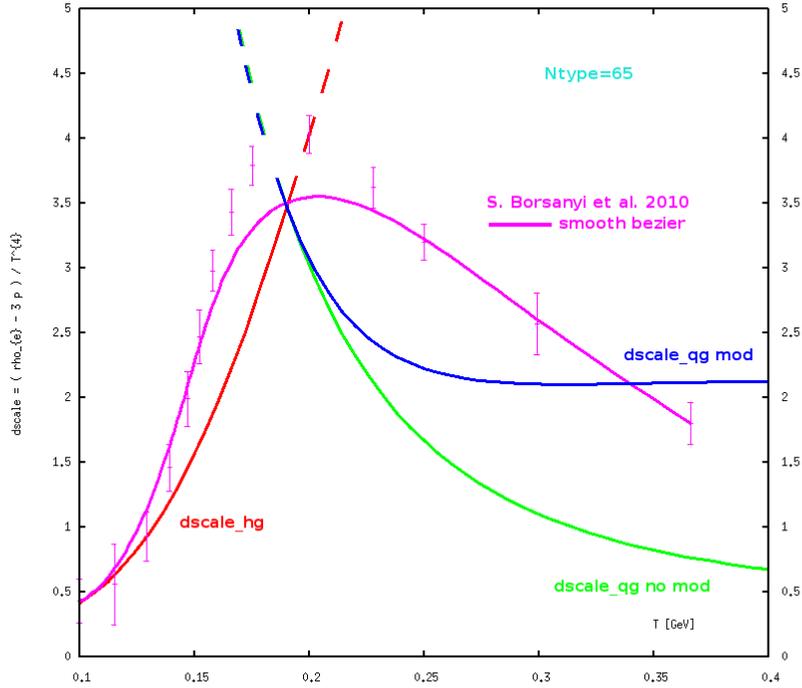}
\vskip -0.4cm
\caption{
The ( dominantly ) second order nature of the transition is displayed 
with respect to the
thermal quantity -- dscale -- forming the shape of an 'indian tent' . 
The same quantity as obtained in Ref. 
\protect \cite{Fodoretal} 
from lattice
simulation of QCD under the same thermal conditions is also plotted
for comparison .
The quantity $\ \mbox{dscale}_{\ qg \ mod}^{\ \nu} \ $
is obtained from Eq. 
\protect \ref{th:14b} ,
here with $\ \nu \ = \ 0.975 \ $, the same as in Figs. 4 - 5 .
The three curves forming the 'indian tent' are represented with dashed lines
outside their range of validity.
}
\label{fig6}
\end{figure}
\end{center}

\newpage
\clearpage


\vspace*{-0.2cm}
\begin{center}
\hspace*{0.0cm}
\begin{figure}[htb]
\vskip -0.3cm
\hskip -0.0cm
\includegraphics[angle=0,width=12.0cm]{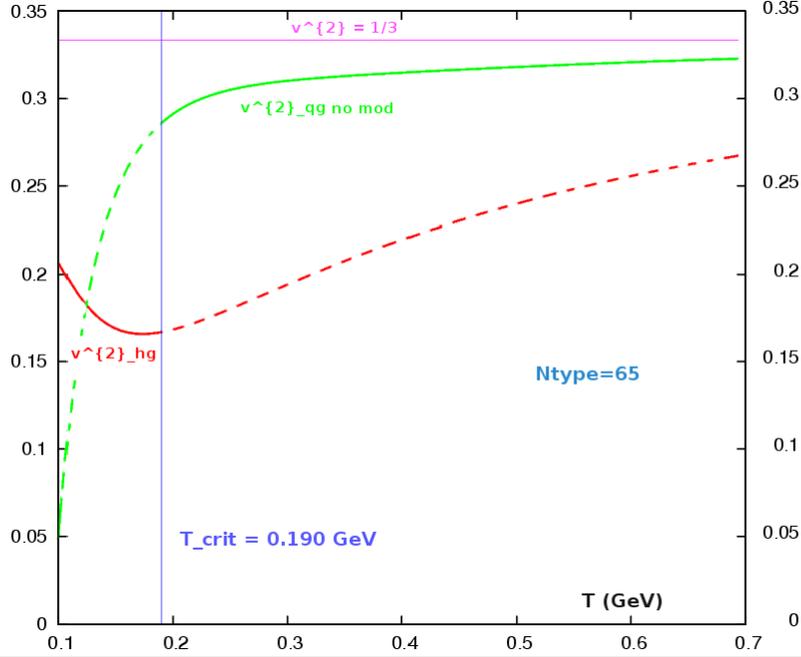}
\vskip -0.2cm
\caption{
The ( dominantly ) first order nature is displayed of the transition 
with respect to the 
square of the velocity of sound , under the same conditions as for
Figs. 4 - 6 , piecewise described by the quantities
$\ v^{\ 2}_{\ hg} \ ; \ \left ( \ hg \ \equiv \ h \ \right ) \ $ for
$\ T \ \leq \ T_{ cr} \ $ from the HRG side and
$\ v^{\ 2}_{\ qg \ no \ mod} $ for
$\ T \ \geq \ T_{ cr} \ $ from the quark-gluon side . 
Only the unmodified setting is used for $\ v^{\ 2}_{\ gq} \ $ .
The two curves for $\ v^{\ 2}_{\ hg} \ , \ v^{\ 2}_{\ gq} \ $ are represented 
with dashed lines outside their range of validity.
}
\label{fig7}
\end{figure}
\end{center}




\newpage

\noindent
The derivations concerning the phase structure of QCD at vanishing
chemical potentials , laid out in the present discussion ,
follow within the
{\it hypothesis} , that the gauge boson pair condensate is at the origin
of stable embeddings of quark flavors -- light and heavy -- into the
( long range ) dynamics controled {\it primarily} by gauge boson 
self-interactions . {\it By the same hypothesis} --
this embedding depends in a minimal way on quark masses , including 
their chiral as well as anti-chiral limits : 
$\ m_{\ \chi_{\ 1}} \ , \ \cdots \ m_{\ \chi_{\ M}} \ \rightarrow \ 0
\ ; \ m_{\ h_{\ 1}} \ , \ \cdots \ m_{\ h_{\ N}} \ \rightarrow \ \infty \ $.
\vspace*{0.05cm}

\noindent
Extensive lattice simulations of the thermodynamic state in particular at zero
chemical potential
have been performed in recent years.
With realistic quark masses for the three light flavours
u,d,s no phase transition but a cross over region is found.
For a review we refer to Ref. [13].

\noindent
For pure gauge theories with gauge group $SUN_c,
N_c>2$ a 1st order phase transition
and for $N_c=2$ a second order transition is found [14,15].

\noindent
The latter phase transition agrees
with the results pertaining to the 3-dimensional Ising model.
\vspace*{-0.2cm}



\subsection*{The argument for -- at least -- a second order phase tran\-sition 
with respect to energy density}

\noindent
In this subsection a short outline is given of how spontaneous parameters 
at $T = 0$ and $T > 0$ are obtained in general renormalizable local field
theories.

\noindent 
We define first in physical space-time, 
the effective action $\Gamma$ for the composite classical field 
$\varphi_{cl} \leftrightarrow \varphi$ associated
with the local gauge boson bilinear, introduced in Eq. \ref{th:2}  

\vspace*{-0.6cm}
\begin{equation}
\label{th:18}
\begin{array}{l}
\exp{\ i W} \ = \ \left \langle \Omega \right |
\exp \left ( i {\displaystyle{\int}} d^{4}x \left ( {\cal{L}} + {\cal{J}} 
\varphi \right ) (x) \right ) \left | \Omega \right \rangle_{ c}
\vspace*{0.1cm} \\
\varphi (x) = \frac{1}{4}  (:) F_{ \mu \nu}^{ A} \ (  x  )
\ F^{ \mu \nu \ A} \ (  x  ) (:)
\hspace*{0.1cm} ; \hspace*{0.1cm} 
{\cal{J}} (x) \ : \ \mbox{source}
\end{array}
\vspace*{-0.0cm}
\end{equation}

\noindent
in the presence of a classical local source ${\cal{J}}$. $_{c}$ in 
Eq. \ref{th:18} denotes the connected part of the generating functionsal $W$.
\noindent
The effective action $\Gamma(\varphi_{c})$ is obtained from
the generating functional $W$ through functional Legendre transform

\vspace*{-0.5cm}
\begin{equation}
\label{th:19}
\begin{array}{l}
\varphi_{cl} (x) = \delta / \delta {\cal{J}}(x) \ W ({\cal{J}})
\hspace*{0.1cm} \rightarrow
\vspace*{0.1cm} \\
\Gamma (\varphi_{cl}) = {\displaystyle{\int}} d^{4}x \ \mbox{{\large (}}
\varphi_{cl} (x) {\cal{J}}(x) \mbox{{\large )}} - W
\hspace*{0.1cm} ; \hspace*{0.1cm}
\varphi_{cl} = \varphi_{cl} ({\cal{J}})
\vspace*{0.1cm} \\
\delta / \delta \varphi_{cl} (x) \ \Gamma = {\cal{J}} (x)
\hspace*{0.1cm} ; \hspace*{0.1cm}
\left . \varphi_{cl} (x) \right |_{{\cal{J}} \rightarrow 0} = {\cal{B}}^{2}
\ , \ \left ( \mbox{Eqs. \ref{th:3},\ref{th:6}} \right )
\vspace*{0.1cm} \\
\Gamma (\varphi_{cl}) = {\displaystyle{\int}} d^{4}x \ v (\varphi_{cl})
\vspace*{-0.5cm}
\end{array}
\vspace*{0.3cm}
\end{equation}

\noindent
The (nonlocal) density functional $v$ defined in Eq. \ref{th:19} 
determines the spontaneous ground state expected value of $\varphi_{cl}$ 
in the limit of vanishing sources as {\it absolute minimum}.

\noindent
It is essential to safeguard appropriate boundary and/or integrability
conditions for the sources ${\cal{J}} (x)$ which trans\-late into
corresponding conditions for the Legendre transforms $\varphi_{cl}(x)$, e.g.

\vspace*{-0.3cm}
\begin{equation}
\label{th:20}
\begin{array}{l}
\lim_{x \rightarrow \infty} {\cal{J}} (x) = 0
\hspace*{0.1cm} \leftrightarrow \hspace*{0.1cm}
{\displaystyle{\int}} d^{4} y \ \left | {\cal{J}} \right |^{2} (y) < \infty
\end{array}
\end{equation}

\noindent
as described in Ref. \cite{MLPM1997} . The functionals $W \ , \ \Gamma$
in Eqs. \ref{th:18}, \ref{th:19} can be extended to Euclidean space.
The absolute thermodynamic limit at zero temperature is reached from
the boundary conditions in Eq. \ref{th:20}
for spontaneous vacuum parameters allowing sources and classical fields 
to approach {\it also} constant values \cite{textbooks}, whereas thermal
environment at finite temperature $T$ corresponds to a finite Euclidean 
time extension

\vspace*{-0.28cm}
\begin{equation}
\label{th:21}
\begin{array}{l}
\Delta \ t_{E} = \beta = 1 / T
\end{array}
\vspace*{0.2cm}
\end{equation}

\noindent
In the finite temperature thermal limit new boundary conditions at 
the bounding times $t_{E} = 0 \ , \ \beta$ 
have to be set, periodic for {\it gauge invariant} bosonic sources and 
classical fields, as considered above relative to the local operator
$\varphi (x)$ defined in Eq. \ref{th:18}. 
Hence we face the two thermodynamic
limits, treated here 
for {\it just} the associated quantities

\vspace*{-0.3cm}
\begin{equation}
\label{th:22}
\begin{array}{l}
\frac{1}{4}  (:) F_{ \mu \nu}^{ A} \ (  x  )
\ F^{ \mu \nu \ A} \ (  x  ) (:) 
 = \varphi (x)
\hspace*{0.1cm} \rightarrow \hspace*{0.1cm}
\left ( \ \varphi_{cl} (x) , {\cal{J}} (x) \ \right )
\vspace*{0.0cm} \\
{\cal{J}} (x) = ( \ 1 / \ g^{2} \ ) - ( \ 1 / \ g^{2} ( x ) \ )
= - \Delta \ \frac{\begin{array}{l} {\scriptstyle 1} \end{array}}
{\vspace*{-0.2cm} \begin{array}{l} {\scriptstyle g^{2}} 
\end{array}} \ ( x )
\vspace*{-0.55cm}
\end{array}
\vspace*{0.3cm}
\end{equation}

\noindent
The external source ${\cal{J}} (x)$ represents a space time dependent
coupling constant, before thermodynamic limits are taken.

\noindent
Absolute thermodynamic ( $T \equiv 0$ ) - and thermal finite temperature 
( $T$ finite ) limits yield the
relations

\vspace*{-0.5cm}
\begin{equation}
\label{th:23}
\begin{array}{c}
\begin{array}{l} {\cal{J}} (x) 
\vspace*{0.1cm} \\
v ( \varphi_{cl} ( x ) )
\end{array}
\hspace*{-0.1cm}
\begin{array}{c} {\scriptstyle \longrightarrow} 
\vspace*{-0.2cm} \\
{\scriptstyle T} \end{array} 
\hspace*{-0.1cm}
\left . \begin{array}{l}  J \in 
\left \lbrack - J_{*} , J_{*} \right \rbrack \rightarrow 0 
\vspace*{0.1cm} \\
v ( \varphi_{cl} , T )
\end{array} \right |
\hspace*{0.1cm} T \hspace*{0.1cm} :
\left \lbrace \begin{array}{l}
\equiv 0
\vspace*{0.1cm} \\
\mbox{finite}
\end{array} \right .
\vspace*{0.2cm} \\
\partial_{\varphi_{cl}} v ( \varphi_{cl} , T ) = J \ \rightarrow \ 0
\ \rightarrow \ \varphi_{cl} = \varphi_{cl} ( T )
\vspace*{-0.5cm}
\end{array}
\vspace*{0.5cm}
\end{equation}

\noindent
Performing the sequence of limits and letting the (constant) values
of the source $J$ vary in a suitable interval 
$\left \lbrack - J_{*} , J_{*} \right \rbrack$ before relaxing it to zero,
as shown in Eq. \ref{th:23}, defines the hysteresis line of the
spontaneous parameter $\varphi_{cl} (J \rightarrow 0, T)$. The latter emerges 
as thermal average for $T$ finite and as vacuum expected value for 
$T \equiv 0$ (Eqs. \ref{th:18}-\ref{th:20}).

\noindent
In the $T$ finite case we have the choice to include chemical potentials ,
one each for conserved quark flavor neutral currents, which we set to zero
here. 

\noindent
The phase transition underlying the present discussion is associated
-- by hypothesis -- with the vanishinging in a singular way at finite critical 
temperature, $T = T_{cr} > 0$ of the quantity $\varphi_{cl} ( T )$ defined in
Eq. \ref{th:23}

\vspace*{-0.2cm}
\begin{equation}
\label{th:24}
\begin{array}{c}
\varphi_{cl} ( T ) \rightarrow 0 \ \mbox{for} \ T \ \rightarrow T_{cr} 
\vspace*{-0.0cm}
\end{array}
\vspace*{0.0cm}
\end{equation}

\noindent
Classical configurations leading to this behaviour inherit a 'Watt-less' nature
from beeing particular to the ground state of QCD \cite{PM1981,HL1981} and thus
should not generate a step-like behaviour of thermal {\it energy density}.
This is borne out in the subtraction of the ground state projection of
the associated energy momentum density tensor $\vartheta_{\mu \nu}$ as shown
in Eqs. \ref{th:4},\ref{th:5}.

\noindent
This however does not imply that the first derivative of energy density 
changes by a finite amount through the transition, i.e. its corresponding
genuinely second order nature. The singularity could well be characterized 
by critical exponents without relation to a definite step in a given derivative
-- of thermal energy density.
\vspace*{0.1cm}


\noindent
The mechanism of gauge boson pair condensation discussed here is a new
analysis taking its roots in material presented in Refs.
\cite{PM1981}, \cite{PMBech1988} and \cite{SKPM2001,SK2001} in chronological 
order.
It is centered on the thermal average of the field strength bilinear
local density operator defined in Eq. \ref{th:2}

\vspace*{-0.2cm}
\begin{equation}
\label{th:25}
\begin{array}{c}
\frac{1}{4}  (:) F_{ \mu \nu}^{ A} \ (  x  ) \ F^{ \mu \nu \ A} \ (  x  ) (:)
\vspace*{-0.0cm}
\end{array}
\vspace*{0.0cm}
\end{equation}

\noindent
The nonzero vacuum expected value ( Eq. \ref{th:3} ) signals a connection
between the localization of color and the structure of Bogoliubov 
transformations as appropriate for gauge fields at long range. 

\noindent
For QCD, actual calculations of effective potentials at large and intermediary
range are not accessible to perturbation theory. Hence  analytic control is
lacking at present.

\section{Conclusions}

\noindent
Our modeling of strong interactions in the qg-phase, in the vicinity but above
the critical temperature $T_{cr} \ \sim \ 0.2 \ \mbox{Mev}$, as described
between Eqs. \ref{th:10} and \ref{th:17}, is supported by observations
of strong coupling in central Au-Au collisions at 200 GeV  at RHIC
\cite{sRHIC,shu}. 
The reduction of pressure and energy density relative to
noninteracting (anti-) quarks and gauge bosons then allows for the
phase transition to be (essentially) of second order with respect to energy
density, as demonstrated in the related thermal quantities shown in
Figs. 2 and 3. 
The square of the velocity of
sound reveals most directly the order of the transition 
through its discontinuity as shown in Fig. 4.

\noindent
The upcoming lower c.m. energy scan at RHIC, new results expected from LHC
concerning hadronic physics and more extended studies in lattice QCD promise to
clarify its phase structure, eventually accompanied by theoretical insights.



\section*{Acknowledgements}
\vspace*{-0.1cm}

\noindent
It is a pleasure to thank the TH-division of CERN for its hospitality,
and Anne Perrin as well as Markus Moser, representing the group of computing
coordinators at the ITP in Bern, for their logistics support.
Topical discussions with Martin L\"{u}scher and Uwe-Jens Wiese are gratefully
acknowledged.



\end{document}